\documentclass[12pt]{iopart}
\usepackage{graphicx}
\usepackage{subfigure}
\usepackage{url}
\usepackage{dcolumn}

\begin{document}

\title[]{Investigating the Morphological Categories \\ in the NeuroMorpho Database by Using Superparamagnetic Clustering}

\author{Krissia Zawadzki$^1$, Mauro Miazaki$^1$ and Luciano da F. Costa$^1$$^2$}

\address{$^1$ Institute of Physics at Sao Carlos - University of Sao Paulo, Avenue
Trabalhador Sao Carlense 400, Caixa Postal 369, CEP 13560-970, Sao Carlos,
Sao Paulo, Brazil}

\address{$^2$ National Institute of Science and Technology for Complex Systems, Brazil}

\eads{\mailto{krissia.zawadzki@usp.br}, \mailto{mauro@ursa.ifsc.usp.br}, \mailto{luciano@ifsc.usp.br}}

\begin{abstract}
The continuing neuroscience advances, catalysed by
multidisciplinary collaborations between the biological, computational,
physical and chemical areas, have implied in increasingly more complex
approaches to understand and model the mammals nervous systems.
One particularly important related issue regards the
investigation of the relationship between morphology and function of
neuronal cells, which requires the application of effective means for
their classification, for instance by using multivariated, pattern
recognition and clustering methods.  The current work aims at such a
study while considering a large number of neuronal cells obtained from
the NeuroMorpho database, which is currently the most comprehensive
such a repository. Our approach applies an unsupervised clustering
technique, known as Superparamagnetic Clustering, over a set of
morphological measurements regarding four major neuronal categories.
In particular, we target two important problems: (i) we
investigate the coherence between the obtained clusters and the
original categories; and (ii) we verify for eventual subclusters
inside each of these categories.  We report a good agreement between the
obtained clusters and the original categories, as well as the
identification of a relatively complex structure of subclusters in the
case of the pyramidal neuronal cells.
\end{abstract}

\maketitle

\section{Introduction}

The progress in neuroscience \cite{Kandel2000} aimed at understanding
the complexity of the nervous system has continuously stimulated
interdisciplinary integration between several scientific fields, such
as biology, computer science \cite{Trappenberg2002}, chemistry, and
engineering. Despite the advances in these fields, there are still
many remaining challenges. In this context, studies focused on
morphological characterization and classification of neuronal cells
have contributed substantially to enhance the neuroscientific
knowledge about the relations between neuronal shape and physiology
\cite{Masland2004,Wen2008, Botaa2007}. These discoveries are
invaluable for several related areas, such as compared anatomy,
neurobiology and diagnosis.

The relationship between neuronal shape and function was first
suggested and investigated by Cajal~\cite{Cajal1989}. Subsequently,
the attention from the neuroscientists shifted to electrophysiological
analysis, which focused on neuronal electrical responses to
stimulus. However, evidences about morphophysiological relationships
have accumulated, such as the correspondence between morphological
and physiological classifications of retinal ganglion
cells~\cite{Waessle1981,Waessle1983}. At the same time, advances in
scientific visualization and shape analysis~\cite{Hosking2009} paved
the way to more comprehensive and sophisticated morphological
approaches, giving rise to the area of computational neuromorphometry,
whose aim is to quantify and study geometrical features of
neurons~\cite{Costa1995}.

The onset of the free-content initiatives, such as in software,
artistic works, and research papers, have motivated the creation and
expansion of public databases. Indeed, a free server of data allows
researchers of several fields to have immediate access to the raw
materials they need in their studies. The use of these databases also
facilitate the replication of experiments and results, as anyone can
access the same data~\cite{Dashti1997}. Another important point is the
prevention of data loss, as the servers cater for backup and data
maintenance.

Currently, the largest repository of neuronal cells is
NeuroMorpho~\cite{NeuroMorpho}. It contains not only the complete
geometrical representation of the cells, but also several
morphological measurements and data information, such as cell type,
species, region, staining method, etc.

The cell categories appearing in such databases frequently take
into account both morpho or eletrophysiological properties.  However,
there is no consensus in the classification of neurons, to the extent
that reclassifications of cells have been reported periodically
(e.g.~\cite{Costa1999}).  The difficulties in obtaining a more
definitive taxonomy are, ultimately, a consequence of the incipient
situation of neuromophological research.  To begin with, there is no
established set of geometrical measurements which should be
used~\cite{Costa1995}. In addition, the lack of a representative
number of cells allied to the choice of specific stochastic and
classification methods also tends to yield varying taxonomies.
Another important problem regarding the classification of neuronal
cells is the identification of meaningful subcategories.

In this context, the availability of significant amounts of neurons
with their morphological properties in the NeuroMorpho repository
motivated an underlying and more systematic study of classification
based on morphometric features.  Though several categories of neuronal
cells have been traditionally adopted in the literature ~\cite{Masland2004,
Cook1998,Stone1983,Botaa2007,Costa1995}, as
a consequence of the largely subjective and incomplete methods used
for their definition, it is not clear how homogeneous these classes
are. Therefore, the investigation of the cell distribution within each
of the main classes provides a particularly important
issue. Especially for the more heterogeneous cases, it is possible
that the existing categories are composed of subcategories which have been
overlooked as a consequence of the subjectiveness and coarseness of
the previous applied measurements and classification methods. The
current work sets out to investigate this important issue, namely by
investigating the uniformity and possible presence of subcategories inside
well-established morphological categories.

In this work, we explore the morphological measurements in NeuroMorpho
by using a established classification procedure of statistical physics,
known as Superparamagnetic Clustering (SPC)~\cite{Blatt1996}. This
method is inspired in a natural magnetic phenomenon presented by
materials due to temperature variations. Described by non-homogeneous
Potts model, where spins at the same state are grouped together, this
physical phenomenon can be applied to data clustering
applications. We used a software available in VCCLAB~\cite{Tetko2005a}.

We chose four large categories of neurons according to the cell type-region
classification and clustered them. The results were then compared to
the original classification. The main objective in this work is to
compare the obtained clusters with the original classification in the
repository, checking the agreement between the original categories and
our clustering. Since SPC is an unsupervised method, it will explore
data and find clusters analysing their features without any subjective
judgment. This approach can either confirm the strength of the
established classification or reveal unnoticed subcategories and
problems in classifications.

Our analysis presents interesting results regarding the classification of the 
selected categories as well as their homogeneity, leading to the necessity to investigate 
the possible factors which are responsible to the observed subcategories and suggesting
a new classification based in the inner relations in classes that exhibits this
behaviour.

This article starts by presenting the adopted database, followed by
the morphological concepts relative to the measurements used in the
neuronal characterization. Afterwards, the theoretical method of
SPC and the software used to analyse the NeuroMorpho data are
explained. Next, the results are presented and discussed.

\section{Materials and Methods}

\subsection{NeuroMorpho Database}

NeuroMorpho~\cite{Ascoli2007} is a contributory database
containing information about digitally reconstructed neuronal cells,
collected by laboratories worldwide. Publicly available, this data is
intended for researches working on issues such as studies of neuronal
system complexity, visualization and neuronal modeling.

The maintenance of the database is provided by the Computational
Neuroanatomy Group of the Krasnow Institute for Advanced Study, from George
Mason University. This initiative is part of the Neuroscience
Information Framework project~\cite{Halavi2008}, endorsed by the
Society for Neuroscience, and includes institutes such as Cornell Univ.,
Yale Univ., Stanford Univ., and Univ. of California.

The database has been periodically updated. Currently, it contains
data relative to 5673 neurons (version 4.0, released in 02/16/2010).
We can access the original and standardized cell morphological
reconstructions, as well as several data and properties such as cell type,
species, brain region, animal width and weight, development age, gender,
used reconstruction methods, magnification, date of upload, images
of spatial structures of neurons, and references to the related literature.

In this work, we separated the data according to the cell type and brain region and
selected the most numerous classes from the database. This yelded the following categories:
\begin{itemize}
    \item Pyramidal cells from Hippocampus (Pyr-Hip);
    \item Medium Spiny cells from Basal Forebrain (Spi-Bas);
    \item Ganglion cells from Retina (Gan-Ret);
    \item Uniglomerular Projection neurons from Olfactory Bulb (Uni-Olf).
\end{itemize}

Table~\ref{tab:number_species} presents the distribution of species within each type-region category.
Since this is a public database, all data used in this article is
available and easily accessible through the NeuroMorpho website. The original category names
are the same as found in the database.

\begin{table}[!ht]
    \centering
    \caption{Number of species in each type-region category.}
    \begin{tabular}{|c|c|c|c|c|}
        \hline
        \hline
        Species     & Spi-Bas & Pyr-Hip & Uni-Olf & Gan-Ret \\
        \hline
        \hline
        Rat         & 232     &  209    &   0     &   0     \\
        Mouse       &  1      &   0     &   0     &  181    \\
        Salamander  &  0      &   0     &   0     &   64    \\
        Drosophilla &  0      &   0     &  233    &   0     \\
        \hline
        \hline
    \end{tabular}
    \label{tab:number_species}
\end{table}

\subsection{Measurements}

\begin{figure}[!ht]
	\centering
		\includegraphics[width=0.6\linewidth]{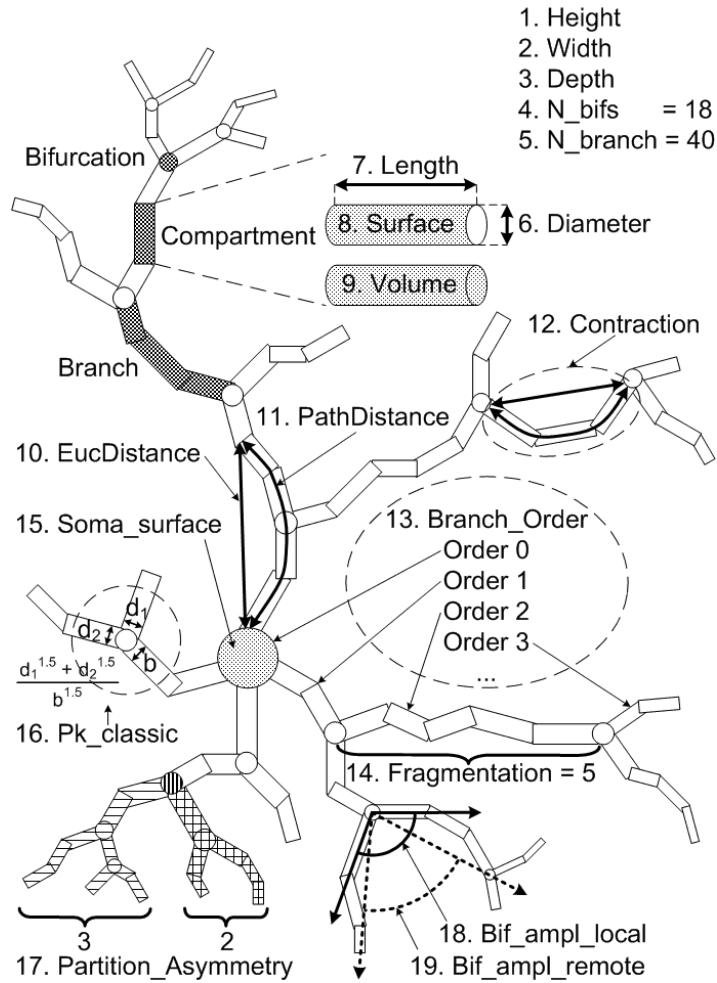}
	\caption{NeuroMorpho measurements.}
	\label{fig:measurements}
\end{figure}

In order to study the morphology of neurons, it is necessary to
represent and characterize them in some way suitable for processing
and analysis. NeuroMorpho provides L-Measure~\cite{Scorcioni2008}, a
tool to extract several measurements from the neurons in the
database. The measurements used in this work are illustrated in
Figure~\ref{fig:measurements}, numbered from 1 to 19 and named as in the
software documentation.

The concepts of compartment, branch and bifurcation are illustrated in
Figure~\ref{fig:measurements}. Compartments are segments represented as
cylinders with diameter and extremity points coordinates. Branches are
formed with one or more compartments between the soma, the
bifurcations and the tips. Bifurcations are points where a branch
splits into two other branches.

Measurements 1, 2 and 3 are the height, width and depth of a neuron,
calculated after its alignment along the principal axis using PCA. The
number of bifurcations and branches in a neuron correspont to the
measurements 4 and 5. The features related to the compartment are from
6 to 9, respectively: diameter, length, surface area and volume.

The branches have their associated measurements numbered from 10 to
14. Measure 10 is the Euclidean distance between a compartment and the
soma, while the path distance (11) is the sum of the lengths of the
compartments between two endpoints.  Contraction (12) is the ratio 
between the Euclidean distance and its path distance. Measure 13 is
the order of the branch regarding the soma, which has order 0. The
branches attached to the soma have order 1. The branches connected to
these branches have order 2, and so on. Fragmentation (14) is the
number of compartments in a branch. Only compartments between
bifurcations or between a bifurcation and a tip are considered.

Measurement 15 is the soma surface area. The soma can be of two types:
a sphere or a set of compartments. In the latter case, the area is
calculated as the sum of the area surfaces of the soma compartments.

The other measurements are related to bifurcations. Pk\_classic (16)
is the ratio $\frac{d_{1}^{r}+d_{2}^{r}}{b^{r}}$, where $r$ is the
Rall's power law value, set in this measure as $1.5$, and $b$, $d_{1}$
and $d_{2}$ are the diameters of the bifurcation compartments (the
parent and the two daughters, respectively). The partition asymmetry
(17) considers the number of tips on the left and on the right
daughter subtrees of a bifurcation as $n1$ and $n2$ in the expression
$\frac{|n1-n2|}{n1+n2-2}$. In Figure~\ref{fig:measurements}, the analysed
bifurcation has vertical stripes, while the left daughter subtree has
horizontal stripes and the right one has a pattern of squares. Then,
in this example, $n1 = 3$ and $n2 = 2$ gives $\frac{|3-2|}{3+2-2} = 0.33$.
Measure 18 is the calculation of the angle between two daughter
compartments in a bifurcation, while measure 19 is the angle regarding
the endpoints of two daughter branches.
Table~\ref{tab:measurement_mean} shows the mean values of the measurements
in each category.

\begin{table}[!ht]
    \centering
    \caption{Mean of the measurements in each type-region category.}
    \begin{tabular}{|c|c|c|c|c|}
        \hline
        \hline
        Measurement       & Spi-Bas & Pyr-Hip   & Uni-Olf  & Gan-Ret \\
        \hline
        \hline
        Height            & 167.43  &  614.88   &  51.78   & 249.11  \\
        Width             & 182.23  &  576.74   &  117.13  & 247.25  \\
        Depth             & 66.94   &  353.68   &  56.54   & 21.01   \\
        N. Bifurcations   & 8.59    &  99.29    &  25.42   & 60.34   \\
        N. Branches       & 24.79   &  209.06   &  52.17   & 131.47  \\
        Diameter          & 0.78    &  0.71     &  1.07    & 0.75    \\
        Length            & 1266.38 &  23654.53 &  499.74  & 4339.40 \\
        Surface           & 2958.64 &  36941.94 &  1764.18 & 9305.34 \\
        Volume            & 999.01  &  9826.63  &  655.88  & 2455.30 \\
        Euclid. Dist.     & 179.08  &  974.86   &  112.58  & 218.86  \\
        Path Dist.        & 224.35  &  2727.95  &  158.07  & 291.82  \\
        Contraction       & 0.86    &  0.85     &  0.92    & 0.86    \\
        Branch Order      & 3.28    &  18.96    &  12.35   & 8.97    \\
        Fragmentation     & 1003.34 &  3311.95  &  446.99  & 2629.73 \\
        Soma Surface      & 715.14  &  1005.75  &  0.00    & 780.84  \\
        Pk Classic        & 1.30    &  1.60     &  1.77    & 1.62    \\
        Part. Asymmetry   & 0.45    &  0.54     &  0.60    & 0.49    \\
        Local Bif. Ampl.  & 63.55   &  69.76    &  93.24   & 82.75   \\
        Remote Bif. Ampl. & 56.30   &  55.24    &  91.62   & 68.52   \\
        \hline
        \hline
    \end{tabular}
    \label{tab:measurement_mean}
\end{table}

\subsection{Superparamagnetic Clustering Method}

The superparamagnetic method is a clustering procedure based on a
physical model of a real material exhibiting magnetic response to some
external parameter. Different from classical approaches, which are restricted to
statistical and mathematical analysis of the system, SPC allows to
evaluate how efficient is the grouping in terms of its intrinsical
properties and has as additional advantages insensitivity to the
initial conditions and robustness to noise. Therefore, it is necessary
to understand the theoretical foundation of the physical concepts
underlying the approach, and how it can be applied to data clustering.

The technique by Blatt, Wiseman and Domany~\cite{Blatt1996} suggests
an analogy with the generalized Ising model, also known as
non-homogeneous Potts model. In this method, the temperature acts as a
parameter controlling the spin configurations of a two-dimesional
atoms network. It is possible to identify three reactions of the
material in response to temperature variations: low, medium and
high temperature intensities. They characterize the respective
ferromagnetic, paramagnetic and superparamagnetic behaviour. The
obtained arrangements of spin orientation is understood as defining
the clustering structure of the data.

The Potts model gives a reference to the energy $E$ of the
configuration, so that stability requires low
energy:
\begin{equation}
\label{eq:hamiltonian}
  \mathcal{H}(s)=-J\sum_{<i,j>}^{N}x_{i}x_{j}
\end{equation}

The transition between the magnetic phases due to variation of
temperature can be characterized by the susceptibility $\chi$, a
physical parameter that reflects the respective overall magnetization
of the system relative to the number of spins with a value between 1
and $q$, calculated as: \begin{equation}
\label{eq:susceptibility}
  \chi=\frac{N}{T}\left(\left\langle m^{2}\right\rangle -\left\langle m\right\rangle ^{2}\right)
\end{equation}
where the variance of magnetization is defined as:
\begin{equation}
\label{eq:magnetization}
  m=\frac{(N_{max}/N)q-1}{q-1}
\end{equation}

The lowest energy probabilistic distribution of the system states
requires predominance of configurations with similar spins for
low temperatures and strong interactions. This distribution is
expressed as:
\begin{equation}
\label{eq:probability}
  P(s)=\frac{1}{Z}\exp\left(\frac{ \mathcal{H}(s)}{T}\right)
\end{equation}
where $Z$ is a normalization constant:
\begin{equation}
\label{eq:normalization}
  Z = \sum_{S} \exp\left(\frac{- \mathcal{H}(s)}{T}\right)
\end{equation}

In order to apply these concepts to data sets and optimize the
execution of the algorithm, the Swendsen-Wang method was adopted. The
motivation of this approach is to analyse different configurations of
the system based on spin neighborhood interactions.

The method proceeds as described in the following. First, we assume a
data set containing $N$ variables $x_i$ whose $d$ components are
measurements of the system features. In analogy to atoms in a
two-dimensional network, we assign a random state $s_i$ among the $q$
possibilities to each respective point $x_i$.

The ranges and steps of temperature variation are predefined in order
to choose the number of interactions in which we want to locate the
superparamagnetic behaviour of system. Then, we analyse the
probability (\ref{eq:probability}) of connection between the sites
with the same spin based on the mutual interaction and neighbourhood
criterion. The latter suggests a maximum number $K$ such that the
interaction $J_{ij}$ between $x_i$ and $x_j$ is computed only if
they are K-nearest neighbors of each other. This interaction is
inversely proportional to the average nearest-neighbor distance $a$
and is mathematically defined as:
\begin{equation}
\label{eq:interaction}
 J_{ij}=\frac{1}{K}\exp\left(-\frac{d_{ij}}{2a}\right)
\end{equation}

By changing the configuration, we estimate the physical parameters of
magnetization (\ref{eq:magnetization}) and the susceptibility
(\ref{eq:susceptibility}). This is repeated until the number $M$ of
interactions is reached.

The threshold temperatures $T_{fs}$ and $T_{ps}$, for which the system is
in the superparamagnetic phase, can be located by identifying the
points of maximal susceptibility and sharp decrease, respectively. In
the transition to paramagnetic regime a guess to $T_{ps}$ can be $T =
\exp^{1/2}/4\ln(1+\sqrt{q})$.

Aimed at quantifying the ordering properties of the new system configuration,
we need to estimate the spin-spin correlation
$G_{ij}=\frac{(q-1)\hat{C_{ij}}+1}{q}$, given in
Equation~\ref{eq:correlation}, which is typically performed by a Monte
Carlo procedure.
\begin{equation}
\label{eq:correlation}
\hat{C}_{ij}={\displaystyle \sum_{\ell=1}^{M}\frac{I_{ij}(\ell)}{M}}\end{equation}
where $I_{ij}=1$ if the points are at the same group or $0$ otherwise.

So, it is necessary to identify the Swendsen-Wang groups with
connected sites, construct the data clusters, compute the magnetization
average $\left\langle \bar{m}\right\rangle =\frac{N_{max}}{N}$ and
then repeat the above procedure until the maximal range of temperature
is reached.

We can verify the different regimes of the superparamagnetic phase
through the susceptibility measure. After locating $T_{fs}$ and
$T_{ps}$, we can analyse the superparamagnetic sub-phases assuming
their mean temperature $T_{clus} = (T_{fs}+T_{ps})/2$ as the point
where the clusters are formed.

The SPC algorithm follows all these steps and concepts, and has many
applications. Currently, there are effective and optimized
implementations available on the web, such as Tetko's program~\cite{Tetko2005}.
We used this software in the current article,
described in the next section.

\subsection{SPC software}

For our data analysis, we used the SPC
software implemented by Tetko~\cite{Tetko2005}. It is developed in
Java, runs on-line as an applet and is freely available through the
VCCLAB (Virtual Computational Chemistry Laboratory) web site at
$<$\url{http://www.vcclab.org/lab/spc}$>$. VCCLAB aims to provide free
on-line tools to analyse chemical data~\cite{Tetko2005a}.

In order to satisfy the required input format, we calculated the
Euclidean distance between all data points and saved them in a text file
to upload into the SPC program. We used all parameters in the
default configuration, since these values are recomended by the
author. The output is analysed and discussed in the following section.

\section{Results and Discussions}

In order to compare the results of SPC and
other approaches of morphological analysis, we applied PCA and LDA
to the set of neuronal cells, which was selected based in the cell type
and brain region. We also verified the agreement between each of the
obtained clusters and the original categories.

\begin{figure}[!ht]
	\centering
		\includegraphics[width=0.6\linewidth]{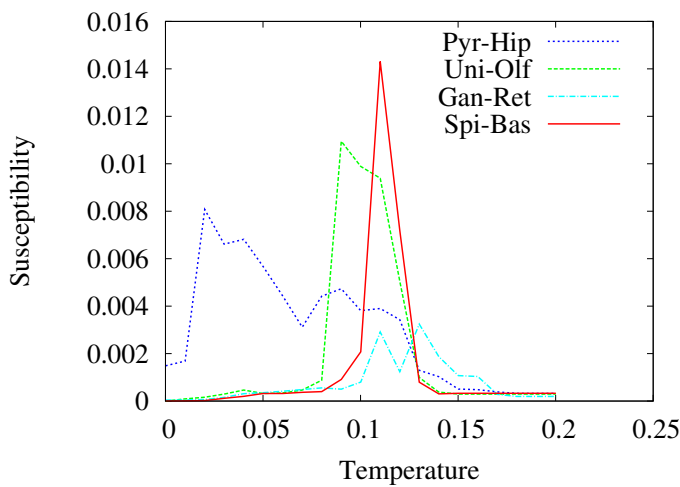}
        \caption{Evolution of the susceptibility parameter to each type-region. Observe the points of maximal and decreasing with which we can identify the superparamagnetic regime.}
	\label{fig:susceptibility}
\end{figure}

\begin{figure}[!ht]
    \centering
    \subfigure[]{
        \label{fig:pca_type_region}
        \includegraphics[width=0.47\linewidth]{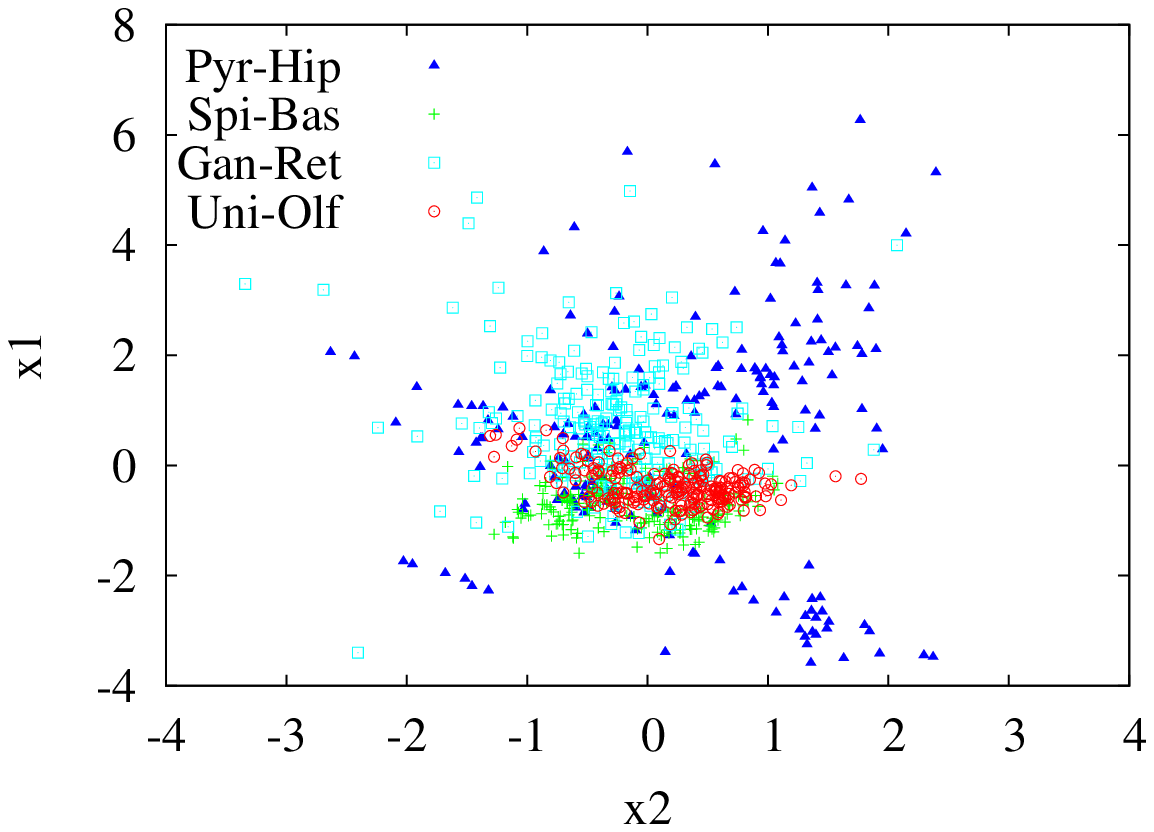}
    }
    \subfigure[]{
        \label{fig:pca_species}
        \includegraphics[width=0.47\linewidth]{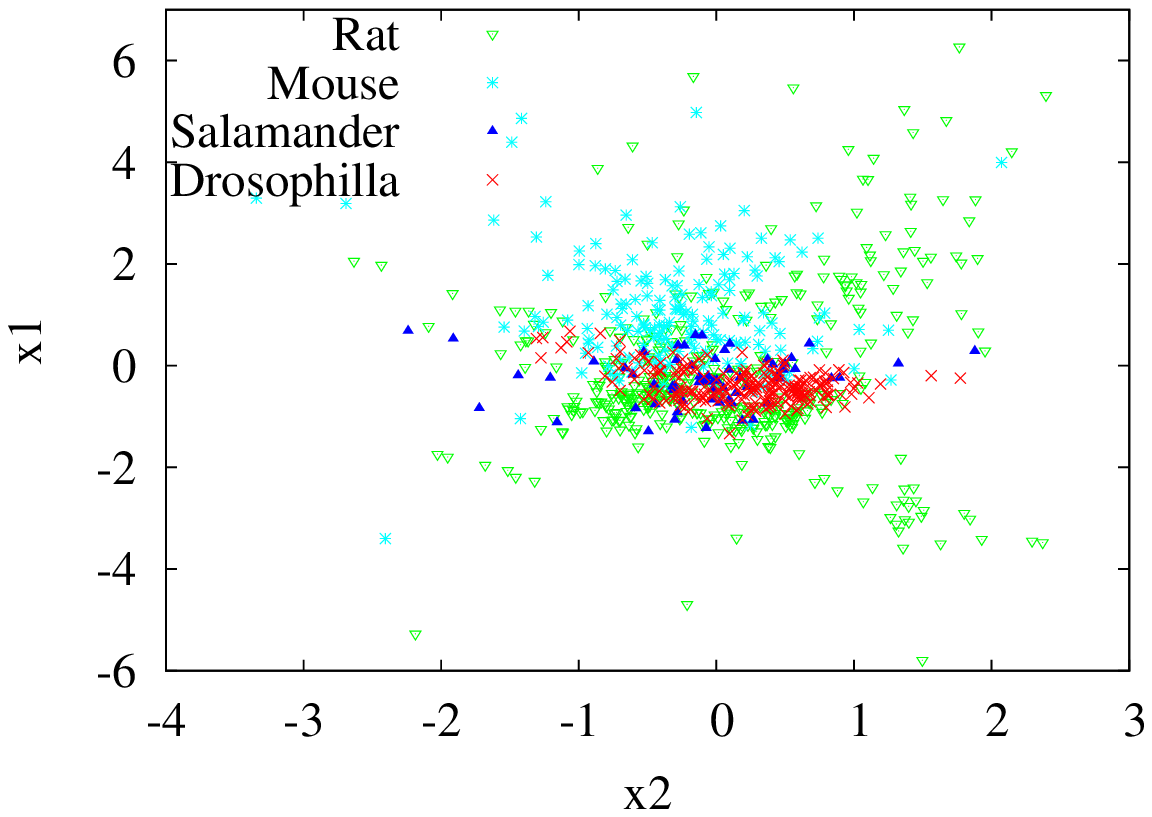}
    }
    \caption{PCA of the four selected cell categories considering the type-region (a) and the species (b) classification. x1 and x2 correspond to the first and second components, respectively.}
    \label{fig:pca}
\end{figure}

\begin{figure}[!ht]
    \centering
    \subfigure[]{
        \label{fig:lda_type_region}
        \includegraphics[width=0.47\linewidth]{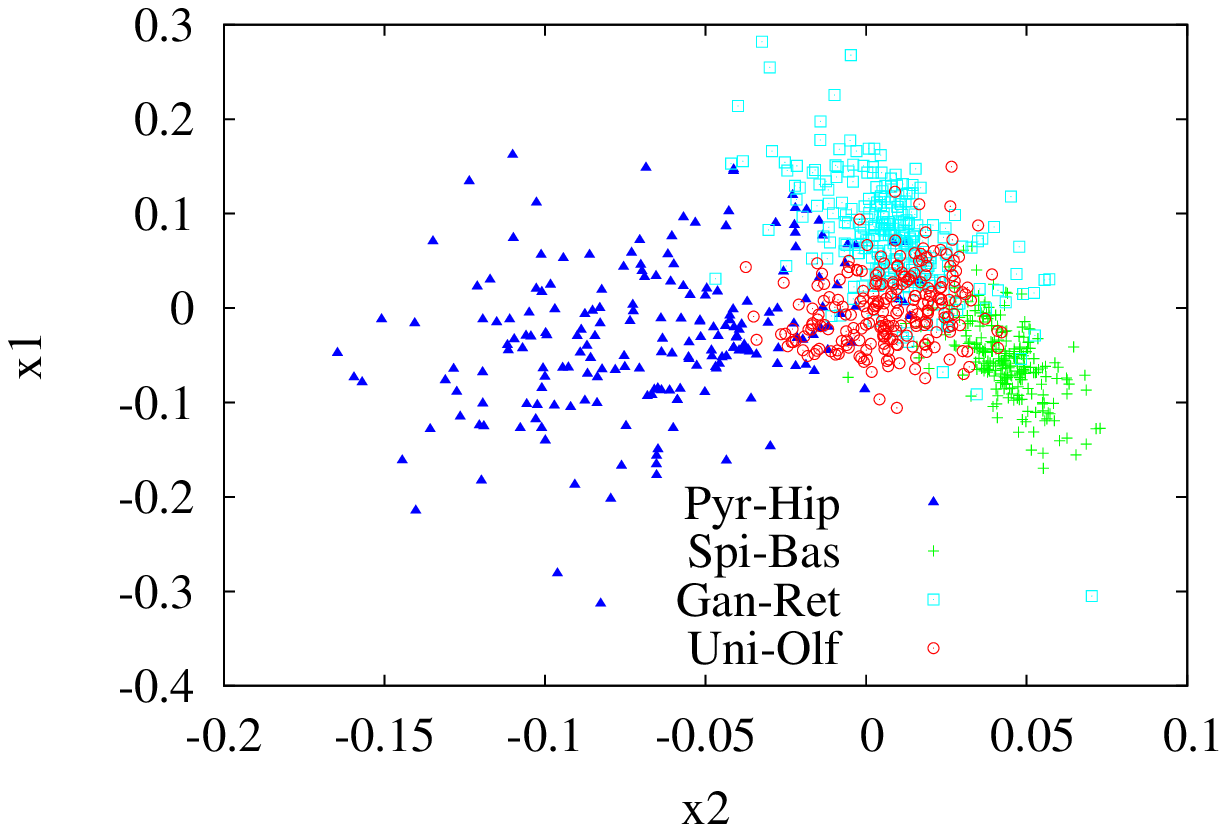}
    }
    \subfigure[]{
        \label{fig:lda_species}
        \includegraphics[width=0.47\linewidth]{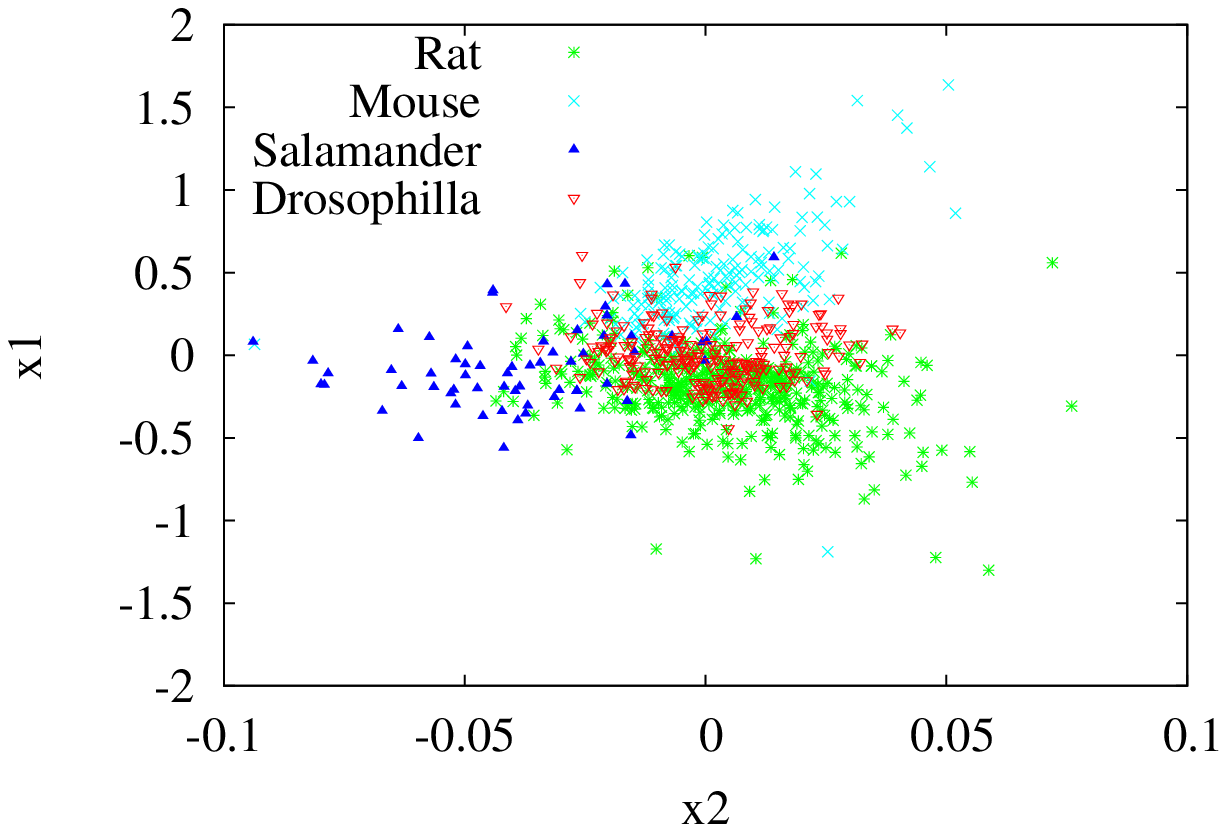}
    }
    \caption{LDA of the four selected cell categories considering the type-region (a) and the species (b) classification. x1 and x2 correspond to the first and second components, respectively.}
    \label{fig:lda}
\end{figure}

In both PCA and LDA, the Pyr-Hip cells are scattered, as we can see
in Figures~\ref{fig:pca_type_region} and \ref{fig:lda_type_region},
respectively. This suggests that these neurons exhibit morphological
features overlapping the other categories, instead of presenting more
homogeneous characteristics which would otherwise imply in their
separation from the other groups. The most homogeneous category is the
Spi-Bas, which appears as a little, compact region in both
methods. The Uni-Olf seems to have a more
central position, intermediate to the other clusters. This behaviour
is also observed for the Gan-Ret, whose location in LDA shows
it to be distinct from the latter category.

\begin{figure}[!ht]
	\centering
		\includegraphics[width=1.0\linewidth]{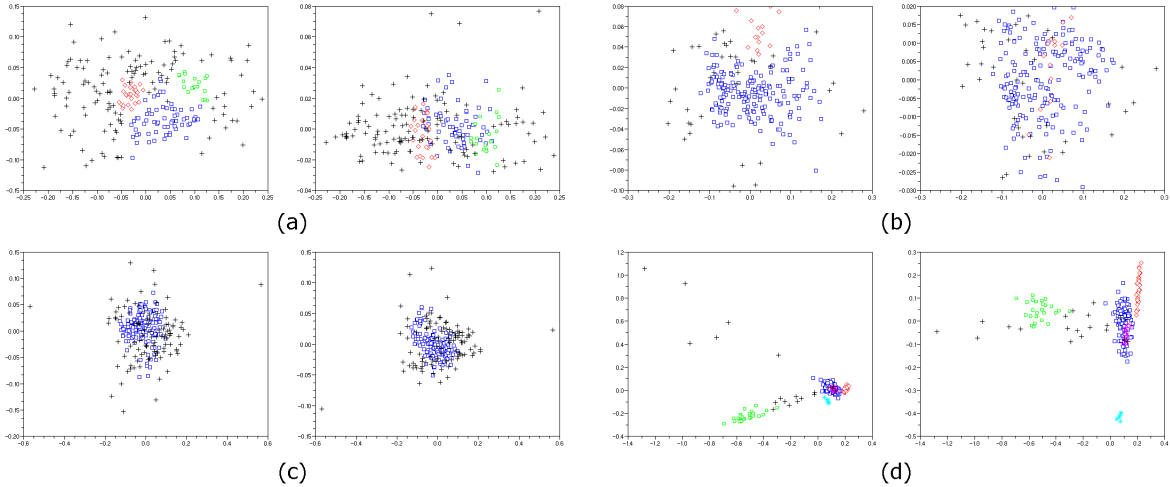}
	\caption{SPC clusters of each category, visualized with PCA: (a) Spi-Bas, (b) Uni-Olf, (c) Gan-Ret, and (d) Pyr-Hip. In each pair of graphics, the left-hand graph shows the components 1 and 2, and the right-hand one exhibits the components 1 and 3.}
	\label{fig:spc_pca}
\end{figure}

\begin{figure}[!ht]
	\centering
		\includegraphics[width=1.0\linewidth]{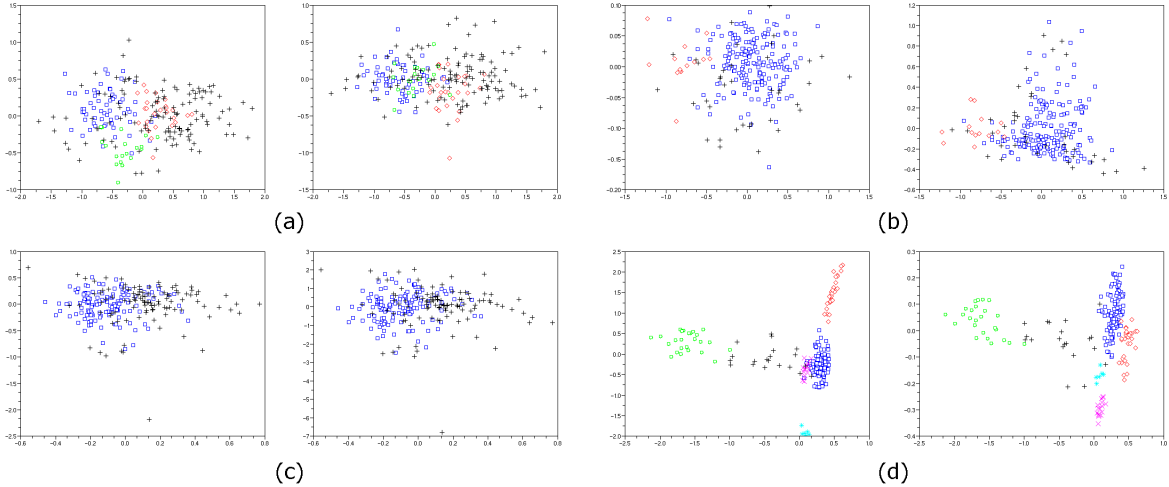}
	\caption{SPC clusters of each category, visualized with LDA: (a) Spi-Bas, (b) Uni-Olf, (c) Gan-Ret, and (d) Pyr-Hip. In each pair of graphics, the left-hand graph shows the components 1 and 2, and the right-hand one exhibits the components 1 and 3.}
	\label{fig:spc_lda}
\end{figure}

In order to analyse the internal composition of the categories, we applied the
SPC method in each one of them and verified the formation of inner
clusters by using PCA (Figure~\ref{fig:spc_pca}) and LDA
(Figure~\ref{fig:spc_lda}) approaches. The Pyr-Hip category revealed
more heterogeneity, splitting in more clusters than the others 
during the SPC process (five clusters at the mean temperature $T_{clus}=0.07$).
This category also presented a
susceptibility curve with a longer superparamagnetic phase,
characterizing a behaviour different from those observed for the other
classes. On the other hand, the Spi-Bas category presented
three clusters not too far from each other and a big and sparse set of
non-clustered cases. In the LDA, the homogeneity is indicated by the
fact that the respective individuals appeared compacted in a specific
region. Regarding the susceptibility curve, its main difference from
the other classes is that it has only one peak at the
superparamagnetic phase.

In the Gan-Ret category, we observed a single cluster in the
superparamagnetic process, and two peaks of susceptibility with
approximate values, relatively lower than the peaks on the other classes. The
resulting clusters agree with the multivariated methods, especially in
a region which overlaps with other categories. We can also see many
non-clustered individuals.

The Uni-Olf category
presented the most characteristic susceptibility curve, giving rise to
two clusters in the obtained distribution. The points are relatively
sparse, and a more numerous subcategory can be seen, which is surrounded by
many unclassified cases and another small cluster.

Another interesting approach regarding the distribution of the
neuronal cells and their categories concerns the investigation about
the relationship between type-region and species, as well as the obtained
clusters. This study was done for both PCA and LDA
(Figures~\ref{fig:pca_species} and \ref{fig:lda_species})
and compared with
the direct references of the data set. We identified the species of
the selected cells, finding them to correspond to rats, mice,
salamanders and drosophillas. So, we count the number of these animals
in each category (see Table~\ref{tab:number_species}) in order to verify
the possible agreement with the results yielded by LDA and PCA.
In most part of the measurements,
it is possible to verify that the Pyr-Hip category has a significative
higher mean value (see Table~\ref{tab:measurement_mean}).

In order to check about possible influences of the original data
properties (e.g. researcher, staining method, etc) on the clusters
obtained in the case of the Pyr-Hip cells, we visualized
the PCA and LDA multivariated projections marked accordingly to these
properties. This category presented five well defined and distinguished clusters
that could indicate influence of the original data properties
or new distinct subcategories.

\begin{figure}[!ht]
	\centering
		\includegraphics[width=1.0\linewidth]{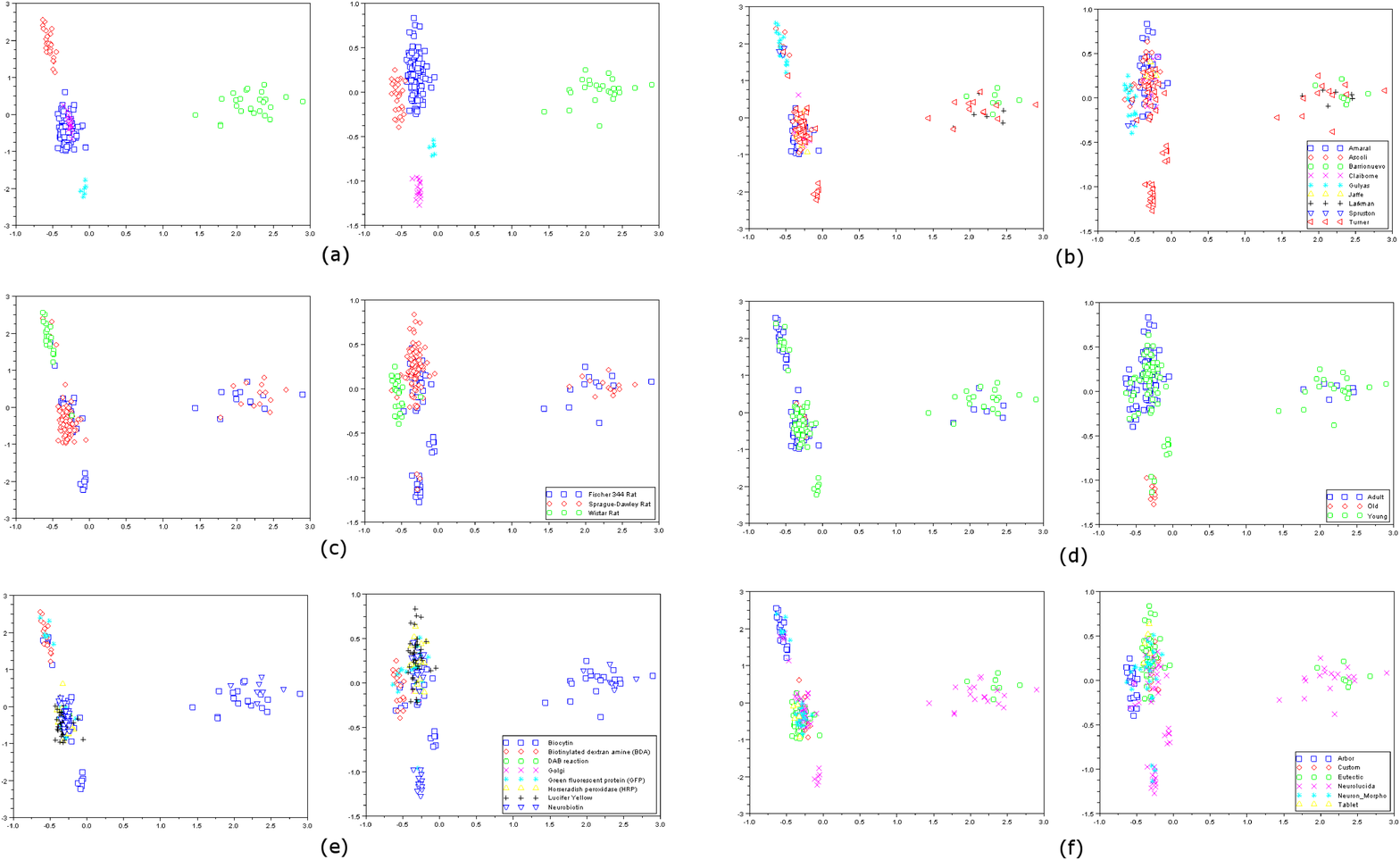}
	\caption{LDA visualization of the clusters found by SPC in the Pyr-Hip category (a), and the original data properties: researcher (b), strain (c), development (d), staining method (e), and reconstruction method (f). In each pair of graphics, the left-hand graph shows the components 1 and 2, and the right-hand one exhibits the components 1 and 3.}
	\label{fig:info}
\end{figure}

The available information about the data is: researcher who provided
the data, animal strain, minimum age, maximum age, age scale, gender,
minimum weight, maximum weight, development, secondary brain region,
tertiary brain region, original format, experimenting protocol, staining
method, slicing direction, slice thickness, objective type, magnification,
reconstruction method, date of deposition and date of upload. In
Figure~\ref{fig:info}, we show only the properties for which we found some
relationship with the clusters. The unclustered elements were eliminated and
the data was reprojected with LDA, in order to obtain a better visualization
of the clusters for this analysis (Figure~\ref{fig:info}(a)).

\begin{table}[!ht]
    \centering
    \caption{Pyr-Hip subclusters size.}
        \begin{tabular}{|c|c|}
            \hline
            \hline
            Cluster      & Number of elements \\
            \hline
            \hline
            Blue square  & 111 \\
            Red diamond  &  29 \\
            Green square &  25 \\
            Magenta x    &  18 \\
            Cyan star    &   6 \\
            \hline
            \hline
        \end{tabular}
    \label{tab:cluster_size}
\end{table}

Table~\ref{tab:cluster_size} shows the number of elements in the clusters of the Pyr-Hip category. The blue square cluster concentrates more than half of the elements, being underlain by a variety of property classes. On the other hand, the small cyan star cluster always presented its elements stable in the same property class through different information properties, but did not distinguish from the others since its classifications were never exclusive.

The green square cluster is formed by the data from researchers (Figure~\ref{fig:info}(b)): Larkman (black cross), Barrionuevo (green squares) and part from Turner (red triangles). This cluster did not lead to a unique and concise classification.

The neuronal data produced by the researcher Gulyas constitute almost all elements in the red diamond cluster and has unique classification in the original data properties: staining method (classified as Biotinylated dextran amine - Figure~\ref{fig:info}(e)) and reconstruction method (classified as Arbor - Figure~\ref{fig:info}(f)). In the strain category (Figure~\ref{fig:info}(c)), the Gulyas data shares its classification with the Spruston data. Both are classified as using the strain 'Wistar Rat'. All but one of the four Spruston elements are in the red diamond cluster.

The magenta x cluster comprises all old rats, as we can see in the development data graphic in Figure~\ref{fig:info}(d). But, it also includes three exceptions (young rats). Examining the data, it was possible to find out that these old rats were 24 months old, while the young exceptions were 1 day old. Thus, although it gets all old rats, there are exceptions at the other side of the age cluster.

Generally, we found no clear correspondence with any of
those a priori properties. The fact that the obtained clustering
structure could not be clearly explained by the original
data properties suggests that the obtained subclusters are a
consequence of some intrinsic morphological variation among the considered cells,
possibly implying the definition of new categories.

\section{Conclusion}

Among several challenges in neuroscience, the morphophysiological
relationship of the neuronal cells has figured as an enhanced approach
in order to establish the characterization and classification of these
structures. Currently, the development of these studies have taken
advantage of information avaliable in public databases, allowing the
application of many pattern recognition and clustering methods.  The
problem has been continuosly investigated, leading to new techniques to classify
a data set. The theoretical bases of these procedures can be
established in physical real models, as Superparamagnetic Clustering.

Aimed at investigating the clustering structures of the neuronal cells, we extracted
many morphological measurements from NeuroMorpho, which is the largest
repository currently, and compared the results of multivariated
and clustering methods in the most numerous categories, whose analysis was
performed considering all category elements and individual categories. In the first case,
our purpose was to verify for agreement between the original classification
and the categories obtained by the PCA and LDA. Afterwards, we isolated
each selected category in order to locate internal clusters
and the respective information that could explain their organization.
The Pyr-Hip cells seemed to form the most heterogeneous category in both
PCA and LDA results, in which remained sparse, as well as in
the SPC results, where it had the higher number of clusters.
The Gan-Ret category was located as
intermediate among other categories and presented two species, despite the
result of the SPC that revelead a single
cluster surrounded by many unclassified cases.
Given that the subclusters obtained for the Pyr-Hip category could not be
explained by specific features of the original cells, we understand
that they potentially imply in a revision of the current
classification in order to account for possible new types of neuronal
cells.

New approaches can be used to complete these results, expanding the
analysis to more properties, performing correlations and eliminating redundancy
between the measurements, as well as the application of other
clustering methods, such as the hierarchical Ward method, motivating further studies.

\section{Acknowledgments}
Mauro Miazaki thanks FAPESP (07/50988-1) for financial support.
Luciano da F. Costa is grateful to FAPESP (05/00587-5) and CNPq
(301303/06-1) for sponsorship.

\clearpage
\section*{References}

\bibliographystyle{unsrt}
\bibliography{neuromorpho}

\end{document}